\documentclass[aps,pra,twocolumn,showpacs,superscriptaddress,floatfix]{revtex4-1}
\usepackage{amsmath}
\usepackage{amssymb,dsfont}
\usepackage[dvips]{graphicx}
\usepackage[small]{subfigure}
\newcommand{\imu}{\textrm{i}}
\newcommand{\bmgscale}{1.0}

\begin{document}

\title{Generation of entanglement density within a reservoir}

\author{C.~Lazarou}
\affiliation{Department of Physics, Sofia University, James Bourchier 5 Boulevard, 1164 Sofia, Bulgaria} 
\author{B.~M.~Garraway}
\affiliation{Department of Physics and Astronomy, University of Sussex, Falmer, 
Brighton, BN1 9QH United Kingdom}
\author{J. Piilo}
\affiliation{Turku Centre for Quantum Physics, Department of Physics and Astronomy, 
University of Turku, FI-20014 Turun yliopisto, Finland}
\author{S. Maniscalco}
\affiliation{Turku Centre for Quantum Physics, Department of Physics and Astronomy, 
University of Turku, FI-20014 Turun yliopisto, Finland}
\date{\today}

\begin{abstract}
  We study a single two-level atom interacting with a reservoir of modes
  defined by its reservoir structure function. Within this framework we are
  able to define a density of entanglement involving a continuum of
  reservoir modes. The density of entanglement is derived for a system with
  a single excitation by taking a limit of the global entanglement.
  Utilizing the density of entanglement we quantify the
  entanglement between the atom and the modes, and also between the
  reservoir modes themselves.
\end{abstract}

\maketitle

\section{Introduction} \label{intro}
In recent years, entanglement has attracted the attention of many physicists 
working in the area of quantum mechanics \cite{Amico2008,Horodecki2009}. 
This is due to the ongoing research in the area of quantum information \cite{Nielsen},
and also because of the advances made in different experimental disciplines, 
such as in
ion traps \cite{Leibfried2003} and Bose-Einstein condensation \cite{Dalfovo1999,Leggett2001}.
Developments in the field of cavity QED, where experiments in the strong
coupling regime are carried out \cite{Raimond2001,Varcoe2004} provide plenty of motivation
for studying quantum information and entanglement. 
Theoretical studies are also important 
in the context of atom-light interactions inside structured reservoirs \cite{Lambropoulos2000}
such as resonant cavities or photonic band gap materials. The theoretically predicted atom-photon 
bound state could also lead to entanglement and
this can also be linked to another problem: that of atom-laser out-coupling  from 
Bose-Einstein condensates \cite{Nikolopoulos2003,Lazarou2007,Nikolopoulos2008}, where analogous effects
were predicted in the past.  

In a recent work by \textit{Cummings} and \textit{Hu} \cite{CummingsN2008}, a two-level
atom coupled to a large reservoir was considered. Starting with simple models
with a few modes, they generalize to reservoirs with a large number of modes, 
where they examine entanglement between the atom and the reservoir. In their analysis
the reservoir is treated as a collective object, i.e.\ the entanglement 
between the atom and each mode, or between the modes, is not considered.

In the present work we study entanglement between the atom and the reservoir
by means of a different approach. In our model, an atom interacts with a continuum 
of modes at zero temperature \cite{Haroche,Breuer}, 
and entanglement properties between the atom and the modes and also
between the modes are considered. Using global entanglement as a measure of entanglement, we 
derive a pair of distributions that can be interpreted as densities of entanglement in terms
of all the reservoir modes. Both distributions can be calculated
in terms of the spectrum of reservoir excitation. With these two new measures of entanglement
we can study in detail entanglement between the atom and the modes, and also between the modes.

In our analysis, we consider a Lorentzian reservoir and cover different
dynamic regimes. For strong coupling we observe the periodic collapses
and revivals of entanglement between the atom and the reservoir, which
are associated with Rabi oscillations of the atom.
Eventually, all the population leaves the atom and the reservoir becomes
entangled. More precisely, 
two bundles
of modes are excited,
forming a ``Bell-like'' state. The method developed here, in
terms of entanglement distributions, can also be used when considering various types
of structured reservoirs. For example, reservoirs with a density of modes characterized by a
band gap \cite{Garraway1997a,Garraway1997b,Lambropoulos2000} can be treated with this method. 
 
The paper is organized as follows.  In section \ref{sec:2} we introduce
a measure of entanglement, the global entanglement, and apply it to our
system-reservoir states when the number of degrees of freedom is finite.
In section \ref{sec:3_2} we introduce the density of entanglement for
the limit of a bath with an infinite number of modes. This entanglement
measure evolves according to our model system as introduced in section
\ref{sec:1}.  In section \ref{sec:3} we present our key results for the
density of entanglement and we
conclude in section \ref{conclusion}.


\section{Measures of multi-partite entanglement} \label{sec:2}

Identifying and measuring entanglement in multi-partite systems presents
various complications. Apart from the case of a two-qubit system, where
entanglement can be identified both for a pure and a mixed state \cite{Hill1997,Wootters1998}, multi-qubit
entanglement is an open problem and to date several measures of entanglement have been proposed 
\cite{Amico2008,Horodecki2009,Mintert2005a,Barnum2004,Meyer2002,Aktharshenas2005}. 
For the analysis that follows we will be using the global entanglement
\cite{Meyer2002} since this will enable us to deal with many modes (or
qubits).
The physical problem we consider is that of entanglement between an atom and a
reservoir, and in the context of this problem we will also consider the entanglement
between different reservoir modes. To quantify this we will take a discrete
bath of reservoir modes. Since we will assume only a single
excitation in each bath mode, we can treat the bath states 
as a set of qubits for the
purpose of computing the entanglement, and then later take a continuum
limit.
With just one excitation in total in the system, this
excitation may be in the atom, or in the bath, so that the state vector
at all times will have the form
\begin{equation} \label{eq:6}
  \vert\psi(t)\rangle=c_0\vert\textbf{0}\rangle+c_a(t)\vert\psi_a\rangle+\sum_{\lambda}c_\lambda(t)\vert\psi_{\lambda}\rangle,
\end{equation}
where $\vert\textbf{0}\rangle$ is the vacuum state
\begin{equation} \label{eq:7}
  \vert\textbf{0}\rangle=\vert0\rangle_a\otimes\vert0000\cdots000\rangle.
\end{equation}
The other two states $\vert\psi_a\rangle$ and $\vert\psi_\lambda\rangle$,
correspond to the atom being excited 
\begin{equation} \label{eq:8}
  \vert\psi_a\rangle=\vert1\rangle_a\otimes\vert000\cdots000\rangle,
\end{equation}
or the mode $\lambda$ of the reservoir being excited
\begin{equation} \label{eq:9}
  \vert\psi_\lambda\rangle=\vert0\rangle_a\otimes\vert000\cdots01_\lambda0\cdots000\rangle.
\end{equation}

Recently this kind of approach has been utilized for the study of
decoherence and entanglement decay in systems with one or two excitations
(effectively at $T=0$)
\cite{Bellomo07,Bellomo08,Maniscalco08,Wang08,Fussel08,Thanopoulos08,Alamri09,Li09,%
Mazzola09a,Mazzola10b,Zhou09,Ferraro09,Francica09,Xiao09,Zhang09,Mazzola09c,Mazzola09b,%
He10,Zhang10a,Man10,Zhou10,Zhang10b,Ge10,Mazzola10a,Tong10b,Wang10,Tong10a,Fanchini10,%
HamadouIbrahim10,Li10,Xu10}. Some of this work examines the decay of a single excitation in a reservoir
made from a continuum of modes
\cite{Fussel08,Thanopoulos08,Mazzola09b,Tong10a}
and other works
 examine entaglement with two or more qubits, but most still utilize the
decay of a single excitation as considered in the present paper
\cite{Bellomo07,Bellomo08,Maniscalco08,Wang08,Li09,Mazzola09a,Mazzola10b%
,Zhou09,Ferraro09,Francica09,Xiao09,Zhang09,Mazzola09c,Ge10,Tong10b%
,Wang10,Man10,Zhou10,Alamri09,He10,Zhang10a}.

If we started with a system of $N$
qubits in a pure state $\vert\psi\rangle$, the global entanglement is defined as
\begin{equation} \label{eq:23}
  Q(\vert\psi\rangle)=2-\frac{2}{N}\sum_{i=1}^N\textrm{tr}\rho^2_i,
\end{equation}
where $\rho_i$ is the reduced density matrix for the $i$-th qubit. When
$\vert\psi\rangle$ is a pure product state then $Q(\vert\psi\rangle)=0$. If
we take an entangled state for  $\vert\psi\rangle$, such as the GHZ state 
\begin{equation} \label{eq:24}
  \vert GHZ\rangle_N=(\vert000\cdots\rangle+\vert111\cdots\rangle)/\sqrt{2},
\end{equation}
then $\textrm{Q}(\vert\psi\rangle)=1$. This example gives a maximum value for the global
entanglement which is normalized such that $0\le Q(\vert\psi\rangle)\le1$. 
Another, more relevant example is the W-state
\begin{equation} \label{eq:25}
  \vert W\rangle_N=\sum_{j=1}^N\vert0\cdots01_j0\cdots0\rangle/\sqrt{N},
\end{equation}
for which the global entanglement goes to zero as $1/N$ for large $N$
\begin{equation} \label{eq:26}
  \textrm{Q}(\vert\psi\rangle)=4(N-1)/N^2.
\end{equation}

An important property of the global entanglement is that it is equal to a sum
over two-qubit concurrences \cite{Horodecki2009}. More specifically, for pure states
$\vert\psi(t)\rangle$ [Eq.\ (\ref{eq:6})] we find from Eq.\ (\ref{eq:23}), see the appendix
for details, that
\begin{equation} \label{eq:27}
  Q(\vert\psi\rangle)=\frac{2}{N+1}C^2(t),
\end{equation}
where $C^2(t)$ reads
\begin{equation} \label{eq:28}
  C^2(t)=\sum_{\lambda=1}^Nc^2(\rho_{a\lambda})+\sum_{1\le\lambda<\mu\le N}c^2(\rho_{\lambda\mu}).
\end{equation}
The concurrence $c^2(\rho_{a\lambda})$ is that for the two-qubit (reduced) density matrix $\rho_{a\lambda}$ 
\cite{Hill1997,Wootters1998}
\begin{equation} \label{eq:29}
  \rho_{a\lambda}=\textrm{tr}_{\mu\neq\lambda}\left\{\vert\psi(t)\rangle\langle\psi(t)\vert\right\}
\end{equation}
between the atom and the $\lambda$-mode. The quantity $\rho_{\mu\lambda}$ is the corresponding density matrix 
for the modes $\mu$ and $\lambda$:
\begin{equation} \label{eq:30}
  \rho_{\mu\lambda}=\textrm{tr}_{a,\kappa\neq\lambda,\mu}\left\{\vert\psi(t)\rangle\langle\psi(t)\vert\right\}.
\end{equation}
The two-qubit concurrence $c(\rho)$ is equal to \cite{Hill1997,Wootters1998}
\begin{equation} \label{conc}
  c(\rho)=\textrm{max}\{0,\sqrt{\lambda_1}-\sqrt{\lambda_2}-\sqrt{\lambda_3}-\sqrt{\lambda_4}\},
\end{equation}
where $\lambda_j$ are the eigenvalues of the matrix
\begin{equation} \label{R}
  R=\rho(\sigma_y\otimes\sigma_y)\rho^\ast(\sigma_y\otimes\sigma_y),
\end{equation}
in decreasing order, i.e. $\lambda_1>\lambda_2>\lambda_3>\lambda_4$, and 
$\sigma_y$ is the relevant Pauli matrix.

Is is important to note here that $C^2(t)$ is exactly equal to the square of the norm of the concurrence 
vector \cite{Aktharshenas2005}, which is one of the many proposed measures for multi-partite entanglement.
Because of the connection between $C^2(t)$ and the two-qubit concurrence Eq.\ (\ref{eq:28}), we  shall refer
to $C^2(t)$ simply as the concurrence.

For the remainder of this work we will focus only on $C^2(t)$ and its properties since, with the 
exception of the normalization factor $2/(N+1)$ in Eq.\ (\ref{eq:27}), it is equivalent to the global 
entanglement. Furthermore, we will be considering a continuum of reservoir
modes, i.e.\ $N\gg1$. In this limit,
as we see in the following section, one can define a density of entanglement for continuous 
systems.

\section{Density of entanglement} \label{sec:3_2}

In the limit of continuous distribution for the reservoir modes, the
summations in Eq.\ (\ref{eq:28}) can
be converted into integrals over the density of modes $\rho_\lambda$, i.e.\
\begin{equation} \label{eq:2}
  \sum_{\lambda}\rightarrow\int \textrm{d}\omega_\lambda\rho_\lambda.
\end{equation}
If we take this limit, the concurrence $C^2(t)$ can be expressed as the
sum of two separate parts which ultimately involve either atomic
population or reservoir mode populations:
\begin{equation} \label{eq:31}
  \begin{split}
    C^2(t)=&\int_{-\infty}^{\infty}d\omega_\lambda\mathcal{E}_A(\omega_\lambda,t)\\ \\
    &+\int_{-\infty}^{\infty}d\omega_\lambda\int_{-\infty}^{\infty}d\omega_\mu\mathcal{E}_R(\omega_\lambda,\omega_\mu,t).
  \end{split}
\end{equation}
The two distributions $\mathcal{E}_A$ and $\mathcal{E}_R$ will form the
entanglement densities and are defined in
terms of the two qubit concurrences to be
\begin{equation} \label{eq:32}
  \mathcal{E}_A(\omega_{\lambda},t)= c^2(\rho_{a\lambda})\rho(\omega_{\lambda}),  
\end{equation}
and 
\begin{equation} \label{eq:33}
  \mathcal{E}_R(\omega_\lambda,\omega_\mu,t)=\frac{1}{2}c^2(\rho_{\lambda\mu})\rho(\omega_{\lambda})
  \rho(\omega_\mu),
\end{equation}
where $\rho(\omega)$ is the reservoir density of modes. The interpretation for these 
two functions is now very simple.

The distribution $\mathcal{E}_A(\omega_\lambda,t)$ is the density of entanglement between the atom 
and all reservoir modes in the vicinity of mode $\omega_\lambda$ i.e.\,  
$\mathcal{E}_A (\omega_{\lambda},t)\textrm{d}\omega_\lambda$ is the total entanglement between the atom 
and all modes in the frequency interval $\omega_\lambda$ and $\omega_\lambda+\textrm{d}\omega_\lambda$. 
In the same way $\mathcal{E}_R(\omega_\lambda,\omega_\mu,t)\textrm{d}\omega_\mu$ gives the entanglement
between the mode $\omega_\lambda$ and modes in the frequency range $\omega_\mu$ to 
$\omega_\mu+\textrm{d}\omega_\mu$. 

For the remainder of this work, $\mathcal{E}_A$ will be referred as
the atom-mode density of entanglement and to $\mathcal{E}_{R}$ as the
mode-mode density of entanglement. The global entanglement
$Q(\vert\psi\rangle)$, Eq.\ (\ref{eq:27}), can be calculated from
these two densities of entanglement, and thus a number of entanglement
properties can be studied in terms of these two distributions.
To this end it is important to note an interesting property for 
the two distributions $\mathcal{E}_A$ and $\mathcal{E}_R$. One can show that both entanglement
distributions can be written in terms of the spectrum of reservoir excitation \cite{Linington2006}
\begin{equation} \label{eq:34}
  S(\omega_\lambda,t)=\rho(\omega_\lambda)\vert c_{\lambda}(t)\vert^2,
\end{equation}
as
\begin{equation} \label{eq:35}
  \mathcal{E}_A(\omega_\lambda,t)=4\vert c_a(t)\vert^2S(\omega_\lambda,t),
\end{equation}
and
\begin{equation} \label{eq:36}
  \mathcal{E}_R(\omega_\lambda,\omega_\mu,t)=2S(\omega_\lambda,t)S(\omega_\mu,t) \,.
\end{equation}
Thus, both entanglement distributions can be derived from the reservoir excitation
spectrum. Having the definition for the entanglement density we can now
introduce a Hamiltonian and dynamics to consider the time-dependent
properties of entanglement for an atom coupled to a Lorentzian reservoir.


\section{Dynamical model} 
\label{sec:1}

The model system consists of a two-level atom coupled to a reservoir of harmonic
oscillators with annihilation and creation operators $a_\lambda$ and
$a_\lambda^\dagger$ respectively. Within the rotating wave approximation the
Hamiltonian reads $(\hbar=1)$
\begin{equation} \label{eq:1}
  \begin{split}
    H=&\sum_{\lambda} \omega_\lambda a_\lambda^\dagger a_\lambda+\omega_0\vert
    1\rangle_{aa}\langle 1\vert
    \\ \\
    &+\sum_{\lambda}g_\lambda\left(a_\lambda^\dagger
    \vert0\rangle_{aa}\langle1\vert+a_\lambda\vert1\rangle_{aa}\langle0\vert\right),
  \end{split}
\end{equation}
where $g_\lambda$ is the coupling between the mode $\lambda$ and the atomic
transition $\vert1\rangle_a\rightarrow\vert0\rangle_a$. The atomic transition
frequency is $\omega_0$ whereas the $\lambda$-mode frequency is $\omega_\lambda$. 
This model will preserve the assumption of a single excitation which is
built into the system states in Eq.\ \eqref{eq:6}.

For the purposes of the analysis that follows, it is very useful to introduce the
reservoir structure function $D(\omega_\lambda)$ which reflects the properties
of the density of modes \cite{Garraway1997a}. This is defined through
\begin{equation} \label{eq:3}
  \rho_\lambda(g_\lambda)^2=\frac{\Omega^2_0}{2\pi}D(\omega_\lambda),
\end{equation}
and is normalized such that
\begin{equation} \label{eq:4}
  \int_{-\infty}^{\infty}\textrm{d}\omega D(\omega)=2\pi.
\end{equation}
With this normalization a measure of the overall coupling strength is $\Omega_0$ which is given by
\begin{equation} \label{eq:5}
  \Omega_0^2=\sum_{\lambda} (g_\lambda)^2.
\end{equation}
Utilizing these assumptions, and a state vector of the form given by Eq.\ \eqref{eq:6},
the Schr\"odinger equation in an interaction picture yields
\begin{eqnarray}
  \imu\frac{\textrm{d}}{\textrm{d}t}\tilde{c}_a&=&\sum_\lambda
  g_\lambda e^{-\imu\delta_\lambda t}\tilde{c}_\lambda, \label{eq:10} \\
  \imu\frac{\textrm{d}}{\textrm{d}t}\tilde{c}_\lambda&=&g_\lambda
  e^{\imu\delta_\lambda t}\tilde{c}_a, \label{eq:11}
\end{eqnarray}
with the detuning between the atomic transition and the mode $\lambda$ being given
by
\begin{equation} \label{eq:12}
  \delta_\lambda=\omega_\lambda-\omega_0.
\end{equation}
The interaction picture amplitudes $\tilde{c}_a$ and $\tilde{c}_\lambda$ are
connected to $c_a$ and $c_\lambda$ via a time-dependent transformation
\begin{equation} \label{eq:13}
  \tilde{c}_a(t)=e^{\imu\omega_0t}c_a(t),\quad\tilde{c}_\lambda(t)=e^{\imu\omega_\lambda
    t}c_\lambda(t).
\end{equation}

Apart from numerical integration of equations (\ref{eq:10}) and (\ref{eq:11}),
one could use analytical methods to derive the dynamics. Examples of these
are: the resolvent method \cite{Lambropoulos2000}, the
Laplace transform \cite{John1994,Kofman1994} and that of the pseudomodes 
\cite{Garraway1997a,Garraway1997b,Dalton01a,Mazzola09b}. This latter method applies when 
the spectral function is analytic with poles in the lower complex plane. Then equations
(\ref{eq:10}) and (\ref{eq:11}) can be replaced, in the continuum limit, 
by a set of equivalent equations where
the atom now couples to a finite set of fictitious modes. Each of these modes
has a one--to--one correspondence to the poles of $D(\omega)$.

For concreteness
we consider Lorentzian reservoir structure functions of width $\Gamma$ and
peak frequency $\omega_c$, i.e.
\begin{equation} \label{eq:14}
  D(\omega)=\frac{\Gamma}{(\omega-\omega_c)^2+(\Gamma/2)^2}.
\end{equation}
This reservoir function has a simple pole at 
\begin{equation} \label{eq:15}
  z=\omega_c-\imu\Gamma/2,
\end{equation}
and for this $D(\omega)$ equations (\ref{eq:10}) and (\ref{eq:11}) are
equivalent to two new equations
\begin{eqnarray}
 \imu\frac{\textrm{d}}{\textrm{d}t}\tilde{c}_a(t)&=& \Omega_0 \tilde{b}(t),
 \label{eq:16} \\
 \imu\frac{\textrm{d}}{\textrm{d}t}\tilde{b}(t)
  &=&
  (\Delta-\imu\Gamma/2)\tilde{b}(t)+\Omega_0\tilde{c}_a(t), \label{eq:17}
\end{eqnarray}
where the atom-cavity detuning $\Delta=\omega_c-\omega_0$, and
$b(t)=e^{-\imu\omega_0t}\tilde{b}(t) $ is the pseudomode amplitude. This set of equations can be
associated with a master equation, where the physical
interpretation for the pseudomode is that of a leaking cavity mode coupled to
the atomic transition \cite{Garraway1997a}.

\begin{figure}
  \begin{center}
    \includegraphics[scale=\bmgscale]{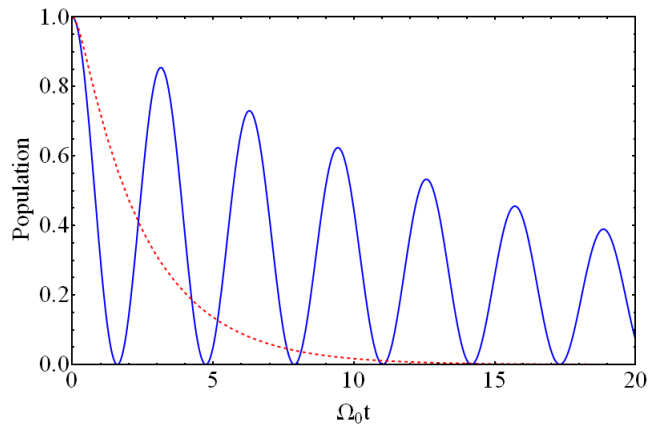}
    \caption{The atomic population $\vert c_a(t)\vert^2$ as a function
      of time for strong coupling $\Omega_0=10\Gamma$ (solid line), and
      weak coupling $\Omega_0=0.1\Gamma$ (dashed line).   Both results were obtained from Eq.\
      (\ref{eq:18}).  The initial system state in this and the following
      figures is defined by $c_a(0)=1$, $c_0=0$ and all $c_\lambda(0)=0$. The detuning $\Delta=0$.}
    \label{fig:1}
  \end{center}
\end{figure}

These coupled ODEs are straightforward to solve, and in particular, for the case of resonance,
$\omega_c=\omega_0$, 
one  finds for the atomic amplitude $c_a(t)$:
\begin{equation} \label{eq:18}
  \tilde{c}_a(t)=\tilde{c}_a(0)e^{-\frac{\Gamma t}{4}}\left(\cosh\left(\alpha t/4\right)+\frac{\Gamma}{\alpha}
  \sinh\left(\alpha t/4\right)\right).
\end{equation}
The modified decay rate $\alpha$ 
is given by
\begin{equation} \label{eq:19}
  \alpha=\sqrt{\Gamma^2-(4\Omega_0)^2}
\end{equation}
in terms of $\Gamma$ and $\Omega_0$.
For strong couplings, $4\Omega_0/\Gamma\gg1$, the atomic population
oscillates between the atom and the reservoir
with a slowly decaying amplitude. In the long time limit all the energy is
lost to the reservoir. If we decrease the coupling $\Omega_0$, the
atom dissipates its energy faster, and for $4\Omega_0/\Gamma\ll1$ the atom exponentially 
decays into the reservoir.
These two types of behavior are illustrated in
Fig.\ \ref{fig:1} which is given as a reference point for the
entanglement discussion in the next section. The figures show 
the atomic population for $\Omega_0=10\Gamma$ 
and $\Omega_0=0.1\Gamma$  plotted against time $t$.

\section{Evolution of reservoir entanglement} \label{sec:3}

\subsection{Entanglement generation by decay} \label{sec:3_1}

\begin{figure}
  \begin{center}
    \subfigure[]{\includegraphics[scale=\bmgscale]{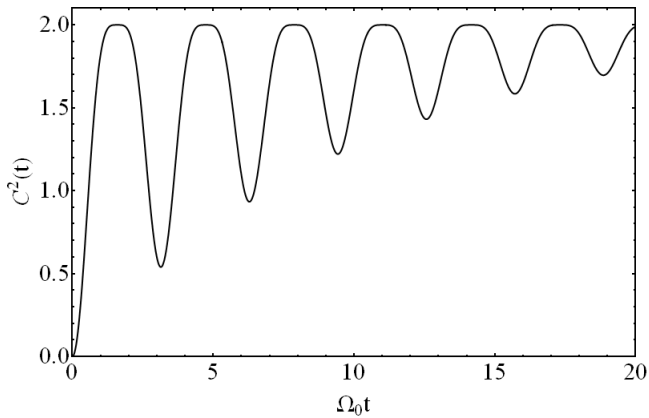}\label{fig:2a}}
    \\
    \subfigure[]{\includegraphics[scale=\bmgscale]{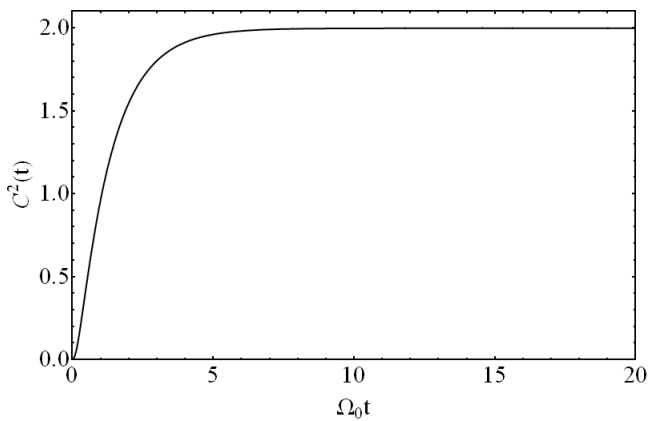}\label{fig:2b}}
    \caption{The concurrence $C^2(t)$ as a
      function of time for different couplings $\Omega_0=10\Gamma$ (a)
      and $\Omega_0=0.1\Gamma$ (b).  The result was obtained from Eqs.\
      (\ref{eq:31}), (\ref{eq:35}) and (\ref{eq:36}).  }
    \label{fig:2}
  \end{center}
\end{figure}

The interaction between the atom and the reservoir results in entanglement creation 
between the atom and the reservoir modes, and also between the reservoir modes. 
This latter entanglement is indirect and is due to the effective coupling between the modes as a result 
of their interaction with the atom. When considering our measure of the total entanglement, 
the concurrence $C^2(t)$ as a function of 
time (Fig.\ \ref{fig:2}), we see that for both strong and weak coupling the concurrence builds up to a
maximum value
equal to $C^2(t)=2$. 
In the case of strong coupling, Fig.\ \ref{fig:2a}, the concurrence
reaches the maximum value very quickly, and then oscillates in a way
which is closely connected to the Rabi oscillations seen in Fig.\
\ref{fig:1}. The oscillations return to the maximum value of two, with
a minimum value that also approaches two as time increases.
If we compare the concurrence to the population in Fig.\ \ref{fig:1}
we see that every minimum in concurrence is matched by a peak in
population of the atomic state. Likewise the maxima in concurrence are
matched by minima in the atomic population: it seems that the decaying
atom is very efficient at generating entanglement in the reservoir.
At this point we note that at long times the energy has left the atom
(see the population Fig.\ \ref{fig:1}) and the system is in an
approximate product state of unexcited atom and bath states. The large
value of the concurrence at these times indicates the presence of
entanglement in the bath.
In the long time limit the concurrence reaches a steady value
$C^2(\infty)=2$, a result that can be derived analytically if one uses
Eqs.\ (\ref{eq:31}) (\ref{eq:35}) and (\ref{eq:36}) and the spectrum
for $t\rightarrow\infty$.
This limit for the concurrence is the same in the weak coupling case,
see Fig.\ \ref{fig:2b}; the atom
decays exponentially and the entanglement reaches a steady state
monotonically.  In the long time limit the atom again disentangles from the
reservoir and entanglement is distributed only between the reservoir
modes.

\begin{figure}
  \begin{center}
    \subfigure[]{\includegraphics[scale=\bmgscale]{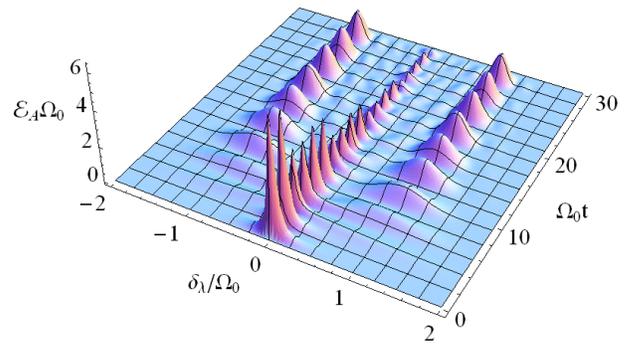}\label{fig:3a}}
    \\
    \subfigure[]{\includegraphics[scale=\bmgscale]{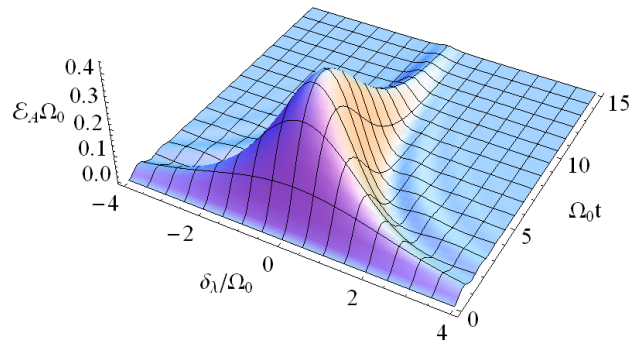}\label{fig:3b}}
    \caption{The density of entanglement $\mathcal{E}_{A}(\omega_\lambda,t)$
      between atom and bath is plotted for different times $t$
      and reservoir mode frequencies $\omega_\lambda$. The coupling strength
      is: (a) $\Omega_0=10\Gamma$; and (b) $\Omega_0=0.1\Gamma$.}
    \label{fig:3}
  \end{center}
\end{figure}

By examining the entanglement densities [Eqs.\ (\ref{eq:32},\ref{eq:33})] we
can gain further insight into where the entanglement resides in this
system and how it evolves over time. In Fig.\ \ref{fig:3} we show the
atom-bath density of entanglement $\mathcal{E}_A(\omega_\lambda,t)$
for both strong and weak couplings. The strong coupling case shows a
complex behavior. First a central peak of entanglement appears in the
vicinity of $\omega_\lambda\approx\omega_0$; note the peak at
$\delta_\lambda\approx 0$ for short times in Fig.\ \ref{fig:3a}.
The entanglement is then transferred to Rabi sidebands at
$\delta_\lambda=\delta_\pm\approx\pm\Omega_0$. The sideband
entanglement oscillations seen take place at half the frequency of the
central peak entanglement oscillations. Ultimately all the
entanglement decays at long times,
$\mathcal{E}_A(\omega_\lambda,t)\rightarrow0$, as there can be no entanglement
between bath and atom when the atomic population approaches zero. At
that point the entanglement indicated by the concurrence in Fig.\ \ref{fig:2a}
must reside in reservoir mode entanglement which is examined in the
next section, Sec.\ \ref{sec:3_3}. Fig.\ \ref{fig:3a} shows
that as Rabi sidebands develop in the reservoir
excitation, the atom-bath entanglement moves from central frequencies
to the sidebands at $\delta_\lambda=\delta_\pm$. The period doubling
of the central peak is due to the oscillations of the excitation
there, combined with the oscillations of atomic population. The
population of the sideband modes is more stable, but oscillations of
the atomic population result in oscillations of entanglement there, too.

For weak coupling, Fig.\ \ref{fig:3b}, the entanglement briefly
resides across the whole reservoir structure. This is essentially
because the coupling of the atom is over this same range. In
weak coupling, however, Rabi sidebands do not develop. Instead the final population
of bath modes will be over a relatively narrow frequency range well within
the model Lorentzian profile, Eq.\ \eqref{eq:14}. Since both atomic, and
mode population is needed for atom-bath entanglement, the narrow
central frequency region is entangled only for a short while
and then decays.

\subsection{Entanglement between reservoir modes} \label{sec:3_3}
As already mentioned, coupling the atom and the reservoir modes will induce an 
indirect coupling between the reservoir modes. Because of this, the modes will entangle and it is
important to consider the properties of the mode-mode density of
entanglement $\mathcal{E}_R$. 
In the long time limit one can
easily obtain analytic expressions for the spectrum of excitation in the
reservoir and from that calculate the density of 
entanglement for $t\rightarrow\infty$.   
Using the definition for the spectrum of reservoir excitation Eq.\ (\ref{eq:34})
with the solution \eqref{eq:18} and Eq.\ \eqref{eq:11}, we find that
for $t=\infty$
\begin{equation} \label{eq:37}
  \begin{split}
    S(\omega_\lambda,\infty)&=
    \frac{\Omega^2_0D(\omega_\lambda)}{2\pi}
    \,\cdot\,
     \frac{\delta^2_\lambda+(\Gamma/2)^2}{(\delta^2_\lambda
    -\Omega^2_0)^2 + (\Gamma/2)^2\delta^2_\lambda}
     \\ \\&
     = \frac{  \Omega^2_0 \Gamma/2 }{ \pi[ 
         (\delta_\lambda^2 - \Omega_0^2)^2   +  (\Gamma/2)^2 \delta_\lambda^2 
     ]},
  \end{split}
\end{equation} 
where, as before,  $\delta_\lambda=\omega_\lambda-\omega_0$ [Eq.\ \eqref{eq:12}].
Taking Eq.\ \eqref{eq:37} together with Eq.\ (\ref{eq:36}) it is straightforward to calculate the mode-mode density of 
entanglement for $t\rightarrow\infty$
\begin{equation} \label{ER}
  \begin{split}
    &\mathcal{E}_R(\omega_\lambda,\omega_\mu,\infty)=\\ \\
    &\frac{\Omega^4_0\Gamma^2}{2\pi^2\left[(\delta^2_\lambda-\Omega^2_0)^2+(\Gamma/2)^2
        \delta^2_\lambda\right]\left[(\delta^2_\mu-\Omega^2_0)^2+(\Gamma/2)^2\delta^2_\mu\right]}.
  \end{split}
\end{equation}

\begin{figure}
  \begin{center}
    \subfigure[]{\includegraphics[scale=\bmgscale]{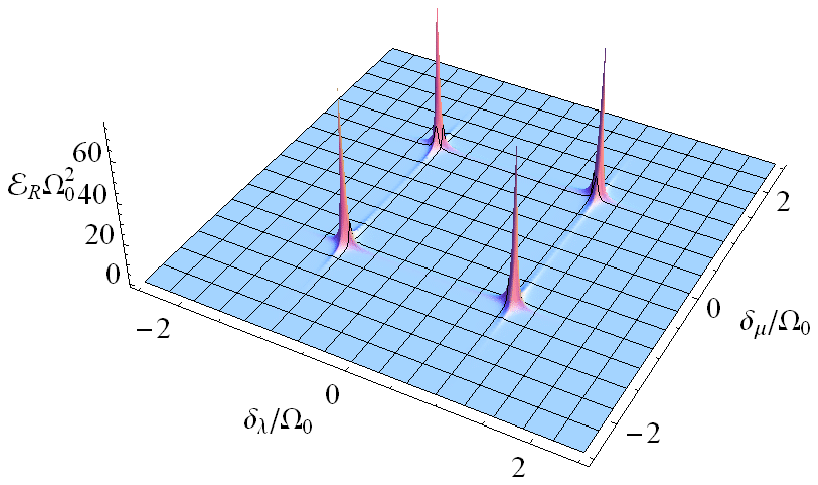}\label{fig:4a}}
    \\
    \subfigure[]{\includegraphics[scale=\bmgscale]{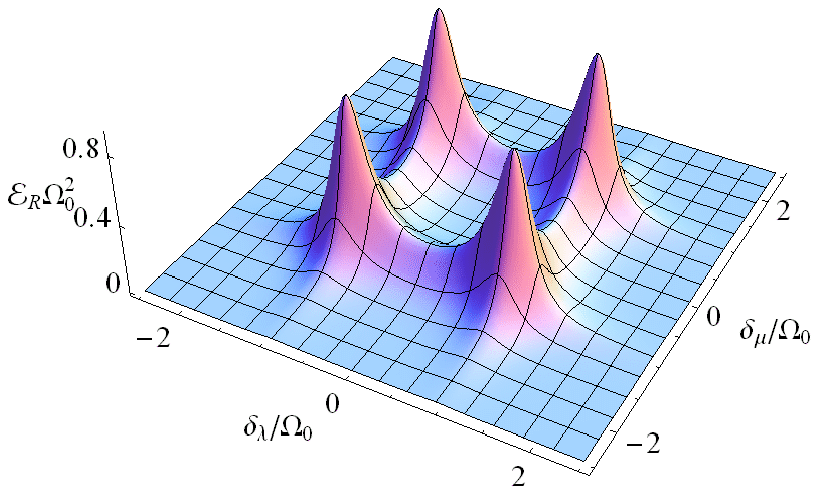}\label{fig:4b}}
    \\
    \subfigure[]{\includegraphics[scale=\bmgscale]{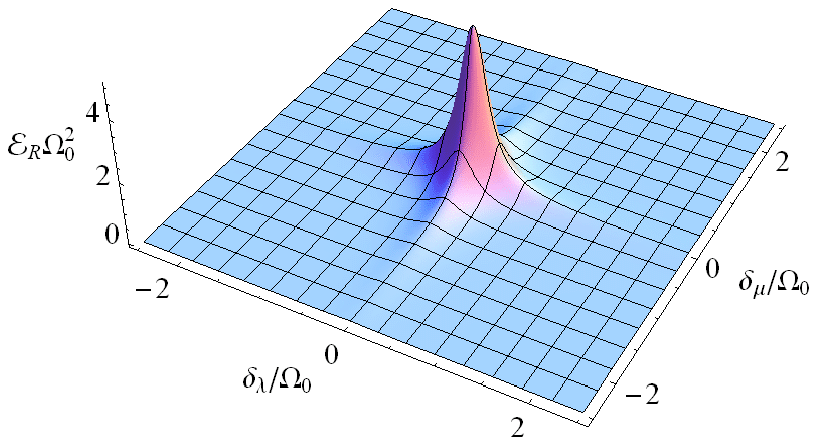}\label{fig:4c}}
    \caption{The density of entanglement
      $\mathcal{E}_{R}(\omega_\lambda,\omega_\mu,t)$ 
      between bath modes 
      for $t\rightarrow\infty$. The coupling strength is:
      (a) $\Omega_0=10\Gamma$; (b) $\Omega_0=\Gamma$; and (c) $\Omega_0=0.1\Gamma$.}
    \label{fig:4}
  \end{center}
\end{figure}

Figure \ref{fig:4} shows the mode-mode density of entanglement for
several couplings and in the limit
$t\rightarrow\infty$. In the strong coupling case, Fig.\ \ref{fig:4a},
we see the formation of four sharp peaks which signify the 
existence of strong entanglement between the two symmetric modes $\delta_\pm$ as a result
of the Rabi splitting.
This picture has a simple and rather intuitive interpretation:
the final state for the reservoir has the form of a Bell-like state
between the two symmetric modes $\delta_{\pm}\approx\pm\Omega_0$.
First we note that the final state takes the approximate form
\begin{equation} \label{eq:38}
  \vert\psi(\infty)\rangle=\vert0\rangle_a\otimes\sum_{\lambda=1}^N
  \left(P_+(\omega_\lambda)+P_-(\omega_\lambda)\right)\vert\psi_\lambda\rangle.
\end{equation}
where
the two probability distributions $\vert P_\pm(\omega_\lambda)\vert^2$ 
are centered at $\delta_\lambda=\delta_+$ and $\delta_-$ respectively. 
For $\Omega_0\gg\Gamma$, i.e.\ in the strong coupling regime, their width is very small which 
practically means that only the two modes $\delta_\pm$ are excited and, for this, the reservoir 
state can be approximately described by a Bell state of the form 
\begin{equation} \label{eq:39}
  \vert\psi(\infty)\rangle\approx\vert0\rangle_a\otimes\vert0_{\omega\neq\omega_0+\delta_\pm}\rangle\otimes
  \left(\vert1_+0_-\rangle+e^{\imu\phi}\vert0_+1_-\rangle\right)/\sqrt{2},
\end{equation}
where $\phi$ is an arbitrary phase factor. This picture applies only at
long times. For short times 
the mode-mode density of entanglement will initially have a distribution that peaks in 
the vicinity of $\omega_\lambda\approx\omega_\mu\approx\omega_0$. As time evolves this distribution 
breaks into four symmetrical peaks seen in Fig.\ \ref{fig:4a}, as a result of the Rabi splitting. 

For moderate and weaker couplings, $\Omega_0 = \Gamma$ and $\Omega_0\ll\Gamma$, the
simple patterns observed in Fig.\ \ref{fig:4a} for the mode-mode
density of entanglement 
due to the Rabi splitting disappear. 
For example in Fig.\ \ref{fig:4b} the density of entanglement for $t\rightarrow\infty$ and 
$\Omega_0=\Gamma$ is plotted. From this we can see that, for moderate couplings, although
the distribution peaks around $\delta_{\mu}=\delta_\lambda=\delta_\pm$, the
entanglement spreads over a wider range of 
reservoir frequencies. For even weaker couplings, Fig.\ \ref{fig:4c}, only modes in
the vicinity of $\omega_\lambda\approx\omega_0$ are excited and thus the entanglement is created only
between modes in this frequency range. 
This is the reason for having a peaked entanglement distribution with a centre at 
$\omega_\mu=\omega_\lambda=\omega_0$.

\section{Conclusion} \label{conclusion} In this work we have presented
a method for studying entanglement between a two-level atom and a
large reservoir of modes such as may be found in, for example, a lossy
cavity field.  Our aim was to study entanglement between the atom and
the reservoir, and within the reservoir itself.
To do this we restricted our interest to a simple case where
the reservoir modes can be treated as qubits. Using global
entanglement, we derived the entanglement densities $\mathcal{E}_A$
and $\mathcal{E}_R$ for the reservoir modes. These distributions have
been given in terms of the well known two-qubit concurrences, and can
be calculated from the reservoir excitation spectrum.  The
entanglement densities are then used both for studying entanglement
between the atom and the reservoir modes, and also between the modes.

In considering different dynamical regimes defined in terms of the
coupling strength, we noticed that when strong interactions occur,
the reservoir modes are entangled in a ``Bell-like'' state 
in the long time limit. This long time limit is a regime where no excitation
remains in the atom and all the entanglement is amongst the reservoir
modes. Since there is no direct interaction between the bath-modes, see
the Hamiltonian \eqref{eq:1}, the final entanglement arises through
indirect interaction. Another way of viewing this is that in the
strong coupling regime we have a non-Markovian system. In such a case,
as the atom decays, some information resides in part of the reservoir
in a way that it can be returned to the atom later \cite{Mazzola09a}. This allows the
indirect coupling between the reservoir modes which creates the
entanglement. In the weak coupling (Markovian) case, the information
does not return to the atomic system and the reservoir entanglement
cannot be created.

In the transition from strong to weak coupling, the  ``Bell-like''
state of Fig.\ \ref{fig:4a} coalesces into a single peaked structure as
was seen in Fig.\ \ref{fig:4c}. Based on the interpretation given above
we would expect that the transition to a single peak takes place as
the system becomes Markovian rather than non-Markovian. In principle
this could be tested with a measure of non-Markovianity \cite{Breuer09}.

In conclusion, when considering entanglement between an atom and a large reservoir, the analysis can be 
formulated in terms of entanglement density functions. These distributions can be associated with 
the spectrum of reservoir excitation, a quantity that can be measured, for example, in cavity QED experiments. 
Although the model considered here, an atom
coupled to a Lorentzian reservoir, is rather simple, it can be extended to consider more complicated
reservoir structures such as model photonic band gaps. Potentially, one could 
consider generalizations for the density of entanglement, and also
extend the model beyond the 
assumption of a single excitation in the system. 

\acknowledgments

This work has been supported in part by the European Commission's ITN
project FASTQUAST. S.M. acknowledges financial support from the Turku
Collegium of Science and Medicine, The Emil Aaltonen foundation and
the Finnish Cultural Foundation. J.P. acknowledges financial support 
from the Magnus Ehrnrooth Foundation and the Academy of Finland 
(project 133682).
 
\appendix*
\section{Global entanglement} \label{append}
For the completeness of our analysis, we show here how the global entanglement 
for the pure state $\vert\psi(t)\rangle$, Eq. (\ref{eq:6}), can be derived. 
First, the state $\vert\psi(t)\rangle$ can be written as a pure state of $N+1$ qubits, i.e.\
\begin{equation} \label{eqA1}
  \vert\psi(t)\rangle=c_0\vert{\bf{0}}\rangle+\sum_{j}c_{j}(t)\vert\psi_j\rangle,
\end{equation}
where $j=a$ for the atom, or $j=\lambda=1,\cdots,N$ for the reservoir modes. The vacuum state 
$\vert{\bf{0}}\rangle$ is given in Eq. (\ref{eq:7}), while $\vert\psi_j\rangle=\vert\psi_a\rangle$ [Eq. (\ref{eq:8})]
for $j=a$, or is given by Eq. (\ref{eq:9}) if $j=\lambda$, i.e.\ refers to a reservoir mode. 

Next we calculate the reduced two-qubit density matrix $\rho_{ji}$ 
\begin{equation} \label{eqA2}
  \rho_{ji}=\textrm{tr}_{k\neq j,i}\{\rho\}
  =\langle\chi_0\vert\rho\vert\chi_0\rangle+\sum_{k\neq j,i}\langle\chi_{k}\vert\rho\vert\chi_{k}\rangle,
\end{equation}
where $\vert\chi_0\rangle$ is the $N-2$-qubit vacuum state
\begin{equation} \label{eqA3}
  \vert\chi_0\rangle=\prod_{k\neq i,j}\vert0_k\rangle,
\end{equation}
and $\vert\chi_{k}\rangle$ is the $N-2$-qubit state with a single excitation, i.e.
\begin{equation} \label{eqA4}
  \vert\chi_{k}\rangle=\vert0\cdots01_k0\cdots0\rangle.
\end{equation}
With these definitions for $\vert\chi_0\rangle$ and $\vert\chi_k\rangle$, the reduced density matrix takes the form
\begin{equation} \label{eqA5}
  \rho_{ji}=\left(\begin{array}{cccc}
    1-\vert c_j\vert^2-\vert c_i\vert^2 &  c^\ast_0c_i &  c^\ast_0c_j & 0 \\
     c_0c^\ast_i & \vert c_i\vert^2 & c_j c^\ast_i & 0 \\
     c_0c^\ast_j & c^\ast_j c_i & \vert c_j\vert^2 & 0 \\
    0 & 0 & 0 & 0
    \end{array}\right),
\end{equation}
where the basis states, starting from top and moving to bottom, are
\begin{equation} \label{eqA6}
  \vert0_j0_i\rangle,\quad\vert0_j1_i\rangle,\quad\vert1_j0_i\rangle,\quad\vert1_j1_i\rangle.
\end{equation}
The concurrence $c(\rho_{ji})$, Eq. (\ref{conc}), for this density matrix reads
\begin{equation} \label{eqA7}
  c^2(\rho_{ji})=4\vert c_{j}\vert^2\vert c_{i}\vert^2.
\end{equation}

In order to calculate the global entanglement, we derive the single qubit density matrix from
Eq. (\ref{eqA5}) by tracing out the $i$-qubit
\begin{equation} \label{eqA8}
  \rho_{j}=\left(\begin{array}{cc}
    1-\vert c_{j}\vert^2 & c_0c^\ast_{j} \\
    c^\ast_0c_{j} & \vert c_{j}\vert^2 \end{array}\right),
\end{equation}
where the basis states are $\vert0_j\rangle$ and $\vert1_j\rangle$. 
Using this we first calculate $\rho^2_j$ and then its trace
\begin{equation} \label{eqA9}
  \textrm{tr}\rho^2_{j}=1+2\vert c_{j}\vert^2\left(\vert c_j\vert^2+\vert c_0\vert^2-1\right),
\end{equation}
which after using probability conservation becomes
\begin{equation} \label{eqA10}
  \textrm{tr}\rho^2_{j}=1-2\vert c_{j}\vert^2\sum_{i\neq j}\vert c_{i}\vert^2.
\end{equation}
Substituting in Eq. (\ref{eq:23}) for the global entanglement we have
\begin{equation} \label{eqA11}
  Q(\vert\psi\rangle)=\frac{2}{N+1}\sum_{j}\sum_{i\neq j}2\vert c_{j}\vert^2\vert c_{i}\vert^2.
\end{equation}
Separating the atom-mode terms and the mode-mode terms, and taking into account 
the symmetry of the two-qubit concurrence
\begin{equation} \label{eqA12}
  c^2(\rho_{ji})=c^2(\rho_{ij}),
\end{equation}
it takes the form
\begin{equation} \label{eqA13}
  Q(\vert\psi\rangle)=\frac{2}{N+1}\left(\sum_{\lambda=1}^{N}c^2(\rho_{a\lambda})+\sum_{1\le\lambda<\mu\le N}c^2(\rho_{\lambda\mu})\right),
\end{equation}
which leads to the result of Eqs. (\ref{eq:27}) and (\ref{eq:28}). 

\bibliography{Manuscript2.bbl}
\end{document}